\documentclass[conference,a4paper]{APSIPA2021}
\usepackage{amsmath}
\usepackage{graphicx}
\usepackage{capt-of}
\usepackage{multirow}
\usepackage{lipsum}
\usepackage{balance}
\usepackage{threeparttable}
\usepackage[backend=biber,style=ieee]{biblatex}
\usepackage{comment}
\usepackage[symbol]{footmisc}
\usepackage[usestackEOL]{stackengine}

\usepackage{nccmath,mathtools}
\usepackage{booktabs, makecell}
\usepackage{float}
\usepackage[caption=false,font=footnotesize]{subfig}
\usepackage{varwidth}
\graphicspath{ {./images/} }
\usepackage{afterpage}
\usepackage{placeins}
\usepackage{cuted} 
\usepackage{geometry}
\usepackage{fancyhdr}

\fancypagestyle{firststyle}{
  \fancyhf{}
  \fancyhead[C]{2023 Asia Pacific Signal and Information Processing Association Annual Summit and Conference (APSIPA ASC)}
}

\begin{document}
\title{After-Fatigue Condition: A Novel Analysis Based on Surface EMG Signals}
\author{
\IEEEauthorblockN {
   Van-Hieu Nguyen$^{1}$,
   Gia Thien Luu$^{2,*}$,
   Thien Van Luong$^{1}$,      
   Mai Xuan Trang$^{1}$
}
\IEEEauthorblockN {
   Philippe RAVIER$^{3}$, 
   Olivier BUTTELLI$^{3}$
}
\IEEEauthorblockA {
  $^{1}$Faculty of Computer Science, Phenikaa University, Hanoi 12116, Vietnam\\
  $^{2}$Biomedical Engineering Department, HUTECH Institute of Engineering, HUTECH University, Ho Chi Minh, Viet Nam\\
  $^{3}$University of Orléans, France PRISME Laboratoire, 12 rue de Blois, BP 6744, 45067 Orléans, France
}
\IEEEauthorblockA {
E-mail: ngvan.hieu613@gmail.com, lg.thien@hutech.edu.vn, \{thien.luongvan, trang.maixuan\}@phenikaa-uni.edu.vn
}
}

\maketitle
\thispagestyle{firststyle}
\pagestyle{fancy}
\begin{abstract}
  This study introduces a novel muscle activation analysis based on surface electromyography (sEMG) signals to assess the muscle's after-fatigue condition. Previous studies have mainly focused on the before-fatigue and fatigue conditions. However, a comprehensive analysis of the after-fatigue condition has been overlooked. The proposed method analyzes muscle fatigue indicators at various maximal voluntary contraction (MVC) levels to compare the before-fatigue, fatigue, and after-fatigue conditions using amplitude-based, spectral-based, and muscle fiber conduction velocity (CV) parameters. In addition, the contraction time of each MVC level is also analyzed with the same indicators. The results show that in the after-fatigue condition, the muscle activation changes significantly in the ways such as higher CV, power spectral density shifting to the right, and longer contraction time until exhaustion compared to the before-fatigue and fatigue conditions. The results can provide a comprehensive and objective evaluation of muscle fatigue and recovery, which can be helpful in clinical diagnosis, rehabilitation, and sports performance. 
\end{abstract}


\pagestyle{empty}
\section{Introduction}
\renewcommand{\thefootnote}{}
\footnote[1]{{*}Corresponding Author}
 Surface electromyography (sEMG) is a non-invasive technique to measure the electrical activity of skeletal muscles. Surface EMG signals can provide valuable information about muscle fatigue, defined as the decrease of maximal force output to maintain or repeat tasks by muscles \cite{enoka_2008_muscle}. Muscle fatigue can affect the performance and health of muscles in various fields such as medicine, sports, rehabilitation, ergonomics, and human-machine interaction \cite{reed_2008_principles,sun_2022_application,basharkatirji_2018_routine}. Therefore, assessing muscle fatigue based on sEMG signals is of great significance and especially interesting to researchers and health professionals. 
 
 Surface EMG signals are a valuable tool to evaluate muscle fatigue, and their applications in life and industrial fields are promising and diverse. However, assessing muscle fatigue based on sEMG is complex because it involves many complex factors, such as muscle type, contraction type, electrode placement, signal processing methods, and fatigue indices \cite{bai_2012_novel}. In addition, surface EMG signals are often affected by noise from various sources, such as motion artifacts, skin perspiration, skin impedance, and electromagnetic interference \cite{zuxiaoqi_2013_evaluation,reaz_2006_techniques}. Therefore, robust and reliable methods are in demand to extract meaningful features from sEMG signals and to classify the states of muscles before, during, and after muscle fatigue. Feature extraction of sEMG signals is a method of extracting useful information from signals under different muscle conditions. Muscle fatigue is usually identified by the decrease of frequency domain indices and time-frequency domain indices and the increase of time domain indices. The traditional approach to acquiring sEMG signals is still commonly used in physiological and clinical studies, based on a pair of electrodes placed on the skin over the muscle. However, the acquired signal depends on the location, the distance between electrodes, the size of the electrode pair, and the area along the muscle fiber, which can lead to very different spectral characteristics and amplification \cite{bai_2012_novel,zuxiaoqi_2013_evaluation}.
 
Initially, sEMG signal features can be divided into three domains: time domain, frequency domain, and time-frequency domain. However, in the past 40 years, a large number of parameters extracted from sEMG signals to evaluate muscle fatigue have been developed. Currently, sEMG features are divided into amplitude-based parameters, spectral-based parameters, non-linear parameters, and muscle fiber conduction velocity (CV) estimation \cite{lloyd_1971_surface,sadoyama_1983_relationships}. For the classical bipolar measurement method, using methods based on more than two electrodes arranged in series allows collecting sEMG signals along the vertical or horizontal axis of the muscle \cite{merletti_2003_the}. A relatively recent approach includes using multiple electrode groups arranged in one- or two-dimensional arrays. In addition, the two-dimensional electrode array can determine the distribution of sEMG amplitude and describe the spectrum over the entire skin area covering the muscle \cite{falla_2007_periodic}. Multi-channel sEMG signals also allow for more accurate and reliable estimation of CV \cite{matteoberettapiccoli_2017_testretest}. Previous studies have investigated the changes in power spectral density (PSD) of sEMG signals during different muscle contraction and load levels. For example, \cite{nagata_1990_emg,lloyd_1971_surface} examined the PSD of sEMG signals at various maximal voluntary contraction (MVC) levels and found that the PSD shifted to lower frequencies with increasing MVC. And \cite{sadoyama_1983_relationships} used a linear array sEMG to study the relationship between CV and the shift of PSD over time with different load levels. They observed that the CV decreased, and the PSD shifted to lower frequencies as the load increased.

Most studies mentioned above have mainly focused on the before-fatigue and fatigue conditions, but the after-fatigue condition has not been well explored. Moreover, previous studies exploited only several separate parameters, which did not fully interpret the objective and comprehensive evaluation of muscle in different conditions.
The main contributions and findings in this work are summarized as follows:
\begin{itemize}
    \item A novel analysis that analyzes all muscle fatigue indicators, such as Root mean square, Mean frequency, Power spectral density, and CV at various maximal voluntary contraction (MVC) levels. In contrast, the previous works analyzed only a few fatigue indicators at a few MVC levels and muscle conditions \cite{troiano_2008_assessment}.
    \item The first attempt to analyze after-fatigue conditions compared to before-fatigue and fatigue conditions. Note that the previous studies only focused on before-fatigue and fatigue conditions as shown in \cite{sadoyama_1983_relationships,nagata_1990_emg,lloyd_1971_surface}. Such analysis and comparison showed a significantly increasing contraction time of the after-fatigue condition until exhaustion, and this phenomenon was partially explained.
\end{itemize}

The rest of this work is constructed as follows. Section~II presents the signal acquisition protocol and our analysis method, and then Section~III discusses the experimental results. Finally, Section~IV concludes the paper.

\section{Materials and methods}
\subsection{Signal acquisition protocol and data pre-processing}
\begin{figure}[htb]
    \centering
    \includegraphics[width=250px]{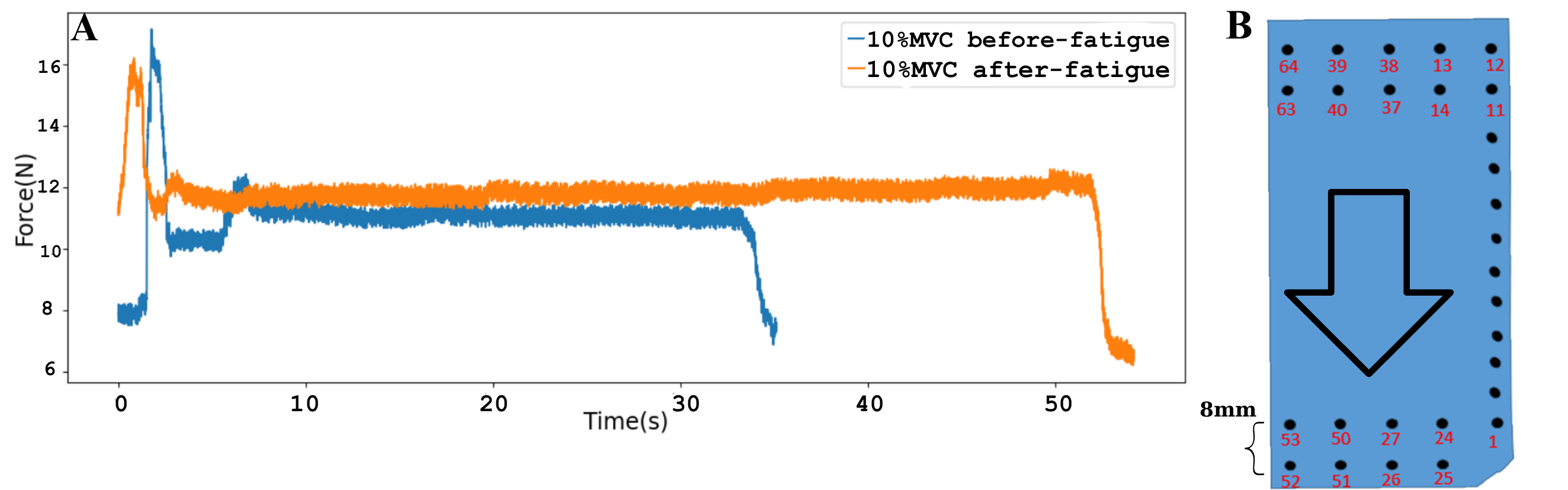}
    \caption{(A) Representation of the longer force's contraction time until exhausted. (B) The matrix sensor is used in signal acquisition.}
    \label{Fig:method}
\end{figure}

A monopolar sEMG signal was collected by the two-dimensional 64-electrode sEMG sensor, divided into thirteen rows and five columns, with an inter-electrode distance of eight millimeters and a sampling frequency of 2048 hertz. The sensor provides five groups of differential signals with thirteen electrodes along with the Biceps Brachii muscle group. A force signal over time is measured from the force gauge collected with each MVC level, as shown in Fig.~\ref{Fig:method} (A). In addition, the representation of the sensor matrix and the direction of signal propagation is described in Fig.~\ref{Fig:method} (B). The sEMG signals from all electrodes obtained from the MVC levels were digitally filtered (20 Hz-400 Hz) to reduce noise.

The data was collected on ten healthy subjects: three women and seven men (mean age 24, standard deviation 1.5 years). The MVC of an individual subject was defined as the average value of three times producing the maximum force, which separated every five minutes. After determining the MVC, the tests can be conducted at different force levels, expressed as a percentage of MVC. The subject should hold the force for at least ten seconds if possible. The experiment starts with a test at 10\% of MVC so the subject can learn to maintain a certain force level.
Furthermore, the process continues with the random order tests at 20\%, 40\%, 60\%, and 90\% of MVC, an exhaustion test at 70\% of MVC, and a final test at 10\% of MVC. Each subject takes a rest of five minutes between two consecutive tests. As such, the states before muscle fatigue include 10\%MVC, 20\%MVC, 40\%MVC, 60\%MVC, and 90\%MVC, while the state during muscle fatigue is 70\%MVC and the state after muscle fatigue is 10\%MVC.

\subsection{Amplitude-based parameters}
Root mean square (RMS) is the main parameter used to investigate the amplitude of the sEMG signal. The formula for RMS is as follows:
\begin{equation}
RMS=\sqrt{\frac{1}{N}\sum_nx^2_n},
\end{equation}
where $x_n$ are the values of the sEMG signal, and $N$ is the number of data samples. RMS is believed that during maximal isometric contractions, amplitude falls progressively, in parallel with the decrease in force \cite{berettapiccoli_2020_evaluation}.

\subsection{Spectral-based parameters}
Two characteristic frequencies that have been used to quantify changes in spectral content based on the Fourier transform are the mean frequency (MNF) and the median frequency (MDF) of the power spectrum
\cite{gonzlezizal_2012_electromyographic}. MDF is calculated as follows:
\begin{equation}
MDF=\int_{f_1}^{f_{median}}PS(f)df =\int_{f_{median}}^{f_{2}}PS(f)df,
\end{equation}
while MNF is calculated as follows:
\begin{equation}
MNF=\displaystyle \frac{\int_{f_1}^{f_2}f.PS(f)df}{\int_{f_1}^{f_2}PS(f)df},
\end{equation}
where $PS(f)$ is the power spectrum calculated from the Fourier transform, $f_1$ and $f_2$ determine the lowest and highest frequency of the signal bandwidth, usually ranging from 20 hertz to 400 hertz. MDF and MNF are related to the change in CV and have been shown in isometric contractions that MNF will shift to lower frequencies during fatigue \cite{merletti_1997_surface}.

\subsection{Conduction velocity estimation}
The muscle's CV was initially calculated by placing electrodes along the muscle.
However, this method can be biased in estimating CV when muscle fibers are not placed in a parallel plane. Moreover, electrodes placed on an undefined muscle domain can lead to misleading information \cite{merletti_2016_surface}.
Recently, multi-channel sEMG signals allowed more accurate estimation of CV both at the global level of the muscle and CV of individual motor units \cite{berettapiccoli_2020_evaluation}. Therefore, this study conducted CV estimation based on the multi-channels maximum likelihood estimation algorithm \cite{mcgill_1984_highresolution}. First, three single differential sEMG signals were selected to remove the mean value. Then, the algorithm calculates two double-differential sEMG signals and tries to maximize the likelihood delay function for observing the delay time between the two signals. CV was obtained after calculating the delay time by dividing the electrode distance by the delay time. The formula for calculating CV is as follows:
\begin{equation}
    CV=\frac{d}{\theta},
\end{equation}
where d is the distance between two electrodes, $\theta$ is the delay time between two electrodes. The algorithm is described in more detail in \cite{mcgill_1984_highresolution}.

CV parameter is related to the properties of the fiber membrane, fiber diameter, and contraction properties of the fiber. Therefore, measuring CV degradation is considered as the strongest indicator of muscle fatigue \cite{fuglevand_1993_impairment}. In addition, changes in the CV in muscle fatigue have a profound impact on the motor unit action potential waveform and, therefore, affect both amplitude parameters and spectral parameters extracted from sEMG, as analyzed in  \cite{gabriel_2009_experimental,lrobinbrody_1991_phinduced}. Previous studies \cite{dimitrova_2003_interpretation,masuda_1983_the}  have also shown that CV is reduced due to the consequences of local metabolic changes in active muscles, mainly due to H+ and K+ distribution in the sarcolemma.

\subsection{The proposed after-fatigue analysis methods}

\begin{figure}[htb!]
    \centering
        \includegraphics[width=0.3\textwidth]{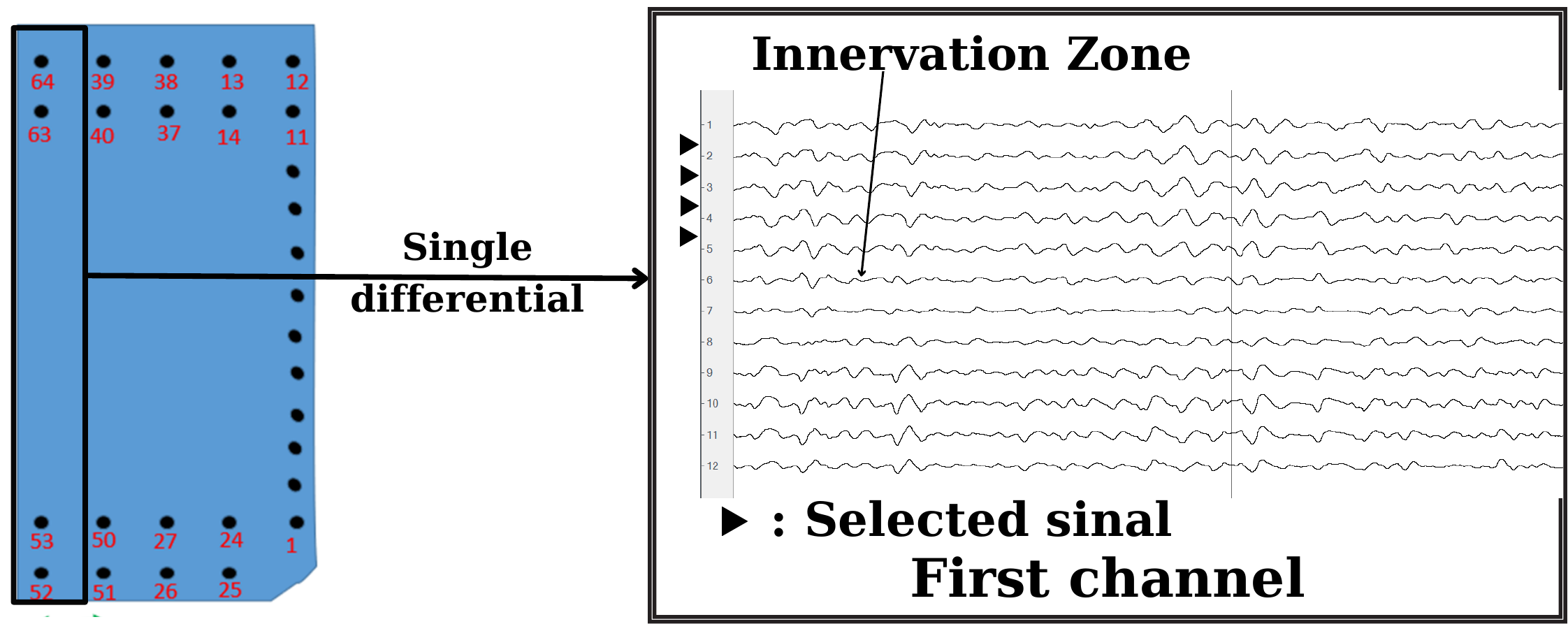}
    \caption{Representation of Innervation zone and signals selection.}
    \label{Fig:method2}
\end{figure}

One of the novel aspects of this analysis method is that it calculates all indicators on the sEMG signal according to MVC levels and contraction time. While in \cite{reed_2008_principles} only shown a few indicators. The amplitude parameters (RMS) and spectral parameters (MNF) were obtained from the entire sEMG signal for each MVC level, excluding the noise parts at the beginning and end. The parameters were averaged over all signals. The electrodes' signals were visualized through a single differential filter to determine the innervation zone, as shown in Fig.~\ref{Fig:method2}.

\begin{figure*}[htb!]
\centering \includegraphics[width=1\linewidth,height=260pt]{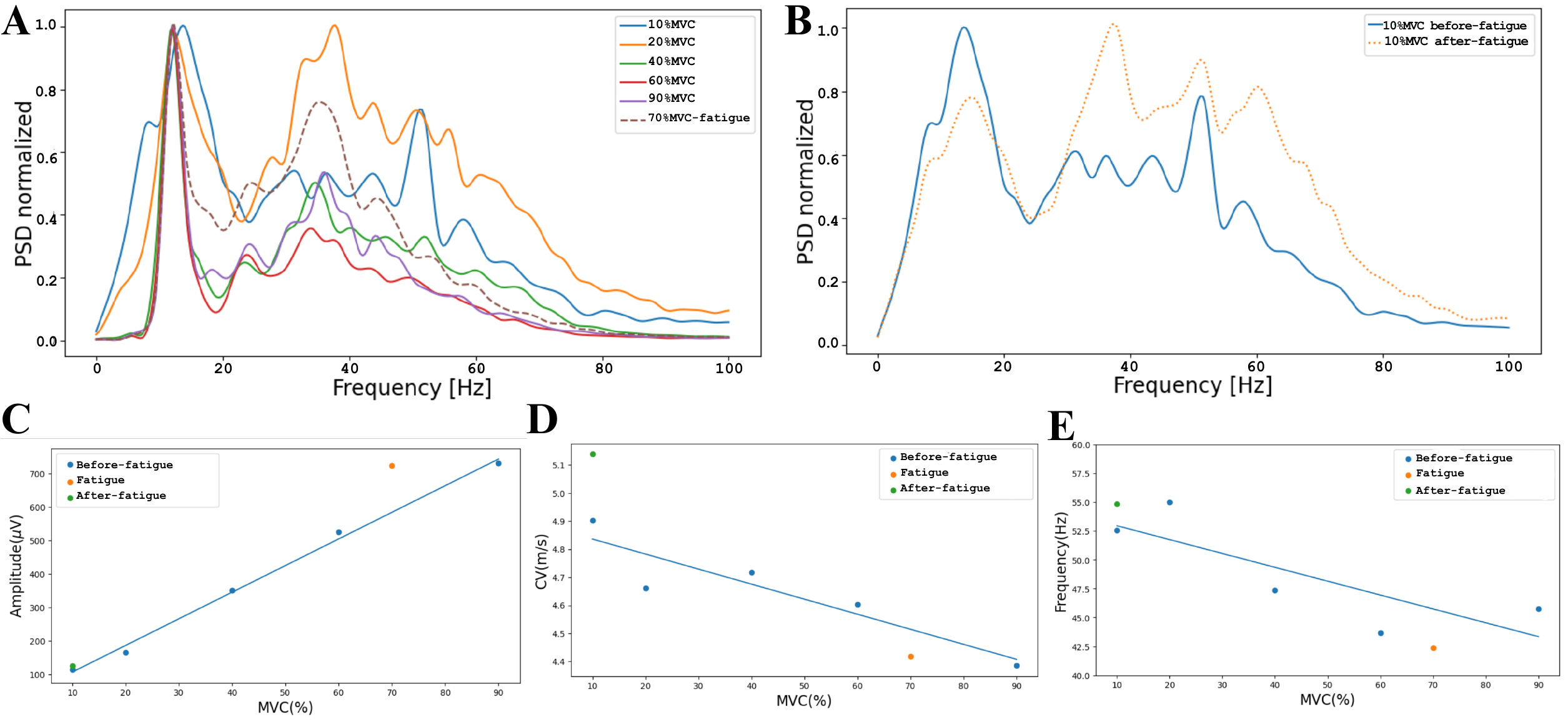}
    \caption{(A) PSD normalized each MVC level. (B) PSD normalized of 10\%MVC before-fatigue and after-fatigue. (C) RMS each MVC level. (D) CV each MVC level. (E) MNF each MVC level.}
\label{Fig:general}
\end{figure*}

The multi-channel algorithm mentioned earlier estimated the CV with selected sEMG signals. The CV value will be accepted if the correlation coefficient between adjacent signals is more than 0.75. The estimated CV value of each MVC level was averaged over channels.

Furthermore, according to the analysis method of contraction time, each selected signal was divided into intervals, spaced 1000ms and 500ms long, which is believed to be stationarity. MNF and RMS obtained from each MVC level were averaged over all selected signals. The PSD representation of the signal was divided into three intervals, each with a length of 1000ms. ``Onset" is taken at the beginning, ``Middle" is taken in the middle, and ``End" is taken at the end of the signal. These intervals allow capturing the changes in sEMG parameters over time and comparing them across MVC levels. 

\begin{figure*}[htb!]
\centering
\includegraphics[width=1\linewidth,height=200pt]{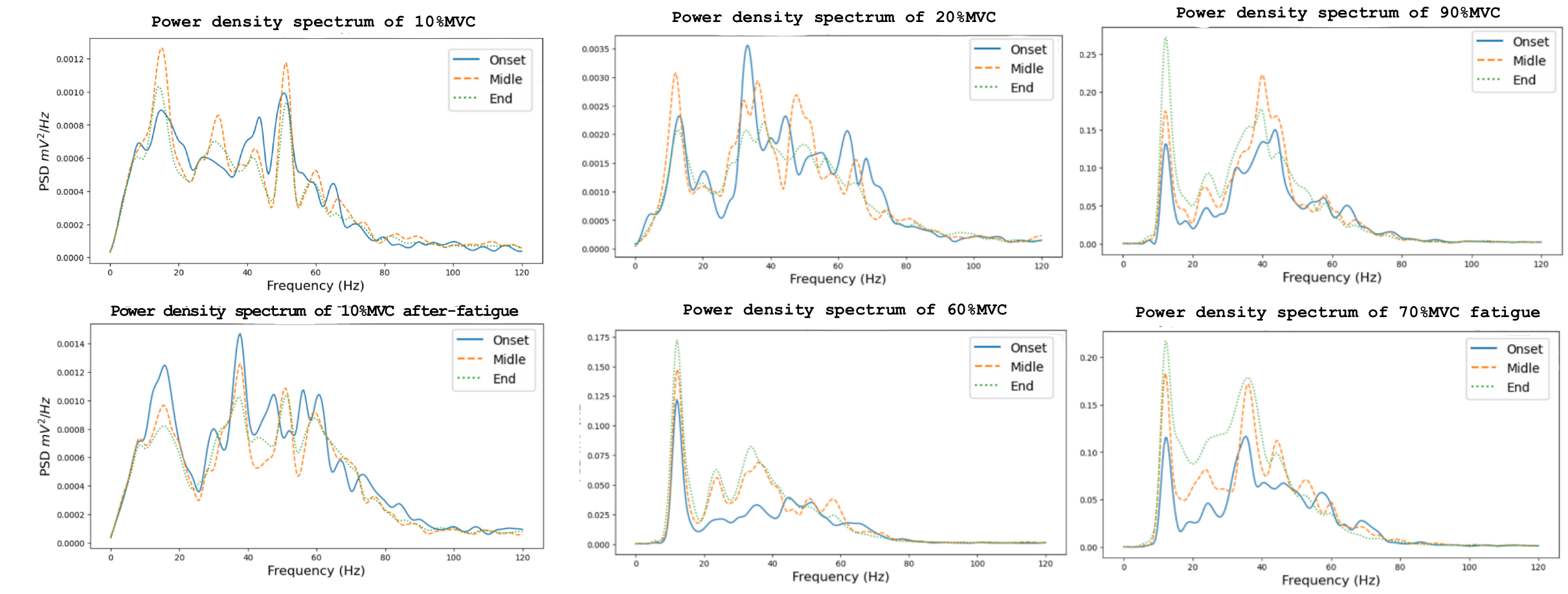}
\caption{Representation of the PSD shift during the contraction time.}
\label{Fig:PSD}
\end{figure*}

Finally, CV was calculated on each selected signal with segments 500ms long and spaced 1000ms apart. The CV value was accepted if the correlation coefficient between adjacent signals was more significant than 0.75. The estimated value of CV for each MVC level was averaged over channels. By using these techniques, the analysis method provides a comprehensive and accurate evaluation of muscle fatigue and recovery.

\section{Experimental Results}
\subsection{Results from analysis of each MVC level}

In this section, The analysis of the PSD, RMS, CV, and MMF of the sEMG signals recorded from the biceps brachii muscle at different levels of maximum voluntary contraction (MVC) was compared with previous studies. The findings' implications according to 10\%MVC after-fatigue are also discussed for muscle fatigue assessment. The results interpretation will be constructed as follows.

The PSD in normalized units for each MVC level is shown in Fig.~\ref{Fig:general} (A). The PSD of 70\%MVC fatigue is more concentrated in the lower frequency range, indicating a higher level of muscle activation. This is consistent with the report by \cite{nagata_1990_emg}. On the other hand, the PSD of other MVC levels tends to shift to lower frequencies as the MVC level increases, suggesting a decrease in muscle activation.

Fig.~\ref{Fig:general} (B) compares the PSD of 10\%MVC before-fatigue and 10\%MVC after-fatigue. A significant difference between the two conditions was first found: the PSD of 10\%MVC after-fatigue has a higher power spectrum in the higher frequency range. In comparison, the PSD of 10\%MVC before-fatigue has a lower power spectrum at the same frequency range. This implies that 10\%MVC after-fatigue signal has a higher muscle activation than 10\%MVC before-fatigue. Note that this phenomenon has not been mentioned in any previous work and needs further analysis at higher MVC levels.

Fig.~\ref{Fig:general} (C) shows the general trend of RMS, which linearly increases to higher MVC levels. This coincides with previous studies such as \cite{gabriel_2009_experimental,dimitrova_2003_interpretation}. The RMS of 10\%MVC after-fatigue and 10\%MVC before-fatigue is almost unchanged. However, the RMS indicator needs to be combined with other indicators, such as MNF and CV, for a comprehensive understanding of muscle fatigue. In addition, 70\%MVC fatigue has a higher RMS amplitude compared to other MVC levels because of higher muscle activation when the muscle is in the fatigue stage, which has been investigated in \cite{merletti_1997_surface,fuglevand_1993_impairment}.

As shown in Fig.~\ref{Fig:general} (D), the general trend of CV linearly decreases to higher MVC levels. This coincides with previous studies such as \cite{sadoyama_1983_relationships}. When the MNF  of 10\%MVC after-fatigue was compared with other MVC levels, we found that the CV of 10\%MVC after-fatigue was the highest compared to other MVC levels. Note that the phenomenon has been overlooked in previous works. Therefore, we suggest that the robust increase in CV affects the PSD of 10\%MVC after-fatigue, as mentioned above. In addition, 70\%MVC fatigue has the lowest CV compared to other MVC because of the muscle fatigue stage.
Finally, Fig.~\ref{Fig:general} (E) represents the general trend of MNF, which linearly increases to higher MVC levels. This also coincides with previous works such as \cite{nagata_1990_emg,gonzlezizal_2012_electromyographic}. The MNF of 10\%MVC after-fatigue was first observed, which has a higher frequency than 10\%MVC before-fatigue; this phenomenon combines with indicators mentioned above that suggested a first evaluation of recovery conditions of muscle. In addition, 70\%MVC has the lowest MNF compared to other MVC levels, which is the same as expected according to the above indicators.

\begin{table*}[hbt!]
\caption{parameter's slope change of each mvc level during contraction time}\label{tab:slope}
\centering
\setlength\tabcolsep{-12pt}
\begin{tabular*}{1\textwidth}[h]{@{\extracolsep{\fill}} l *{10}{c}}
\toprule
Parameters & \multicolumn{9}{c}{ MVC levels} \\
\cmidrule{2-11}
& 10\%MVC & 20\%MVC & 40\%MVC & 60\%MVC & 90\%MVC &
70\%MVC Fatigue & \addstackgap{\stackanchor{10\%MVC}{(After-fatigue)}} \\
\midrule
MNF&0.011&0.016&-0.321&-0.466&-1.019&-0.64&0.007
\\
RMS&-0.357&-0.051&7.987&15.417&22.376&21.093&-0.004\\
CV&-0.001&-0.013&-0.019&-0.026&-0.045&-0.054&0.003\\
\bottomrule
\end{tabular*}
\end{table*}

In summary, novelty findings were found according to 10\%MVC after-fatigue compared to other MVC levels. In addition, a shift of PSD to the right at 10\%MVC after-fatigue was first found, compared to 10\%MVC before-fatigue. This phenomenon comes with a robust increase in MNF and CV compared to other MVC levels. Based on the findings, We suggested a novelty evaluation of the after-fatigue conditions of muscle, where the shift to the right of PSD and increasing MNF denote higher muscle activation and a robust increase in CV denotes the reabsorption and distribution of ions such as H+ and K+ in the muscle membrane, leading to an increase in CV, as mentioned earlier. This further proved the more extended maintenance of contraction time at 10\%MVC after-fatigue conditions.

\subsection{Results from analysis contraction time}

This section represents an analysis of the PSD indicator over time and evaluates the slope change of each parameter, such as MNF, RMS, and CV. Moreover, findings according to 10\%MVC after-fatigue compared to other MVC levels and discuss their implications for muscle fatigue assessment. The results interpretation will be constructed as follows.

Fig.~\ref{Fig:PSD} represents the analysis of the PSD indicator over time. The compression trend of PSD was first generalized at various MVC levels and conditions over time, while \cite{nagata_1990_emg} only exploited PSD on different MVC levels and some conditions. The PSD has a wide frequency, and the mean frequency is almost unchanged according to 10\%MVC and 20\%MVC before-fatigue. However, when MVC levels are increased, PSD tends to compress at a lower frequency apparent over time according to 60\%MVC and 90\%MVC. In addition, when the muscle is in the fatigue stage according to 70\%MVC, the compression of PSD is not only at a lower frequency but also a robust increase in the power spectrum. Note that this finding has been overlooked in the previous studies.

In Table~\ref{tab:slope}, the slope change of all parameters such as MNF, RMS, and CV of each MVC level during contraction time were summarized. Firstly, we observed a general trend of MNF’s slope at before-fatigue MVC levels. We first found that the higher the MVC levels, the more apparent MNF’s slope decreases, which means the muscle tends to fatigue faster as higher force is required. In addition, MNF’s slope of 70\%MVC fatigue seems unaffected in fatigue conditions. MNF’s slope of 10\%MVC after-fatigue is nearly zero, which is almost unchanged over time and needs further analysis to understand fully.

In contrast to the general trend of MNF’s slope, RMS’ slope tends to increase as higher force is required according to before-fatigue conditions, as shown in Table~\ref{tab:slope}. In addition, RMS’s slope of 70\%MVC fatigue is nearly equal to 90\%MVC, which means muscle needs more power to retain force than other MVC levels in fatigue conditions. In 10\%MVC after-fatigue conditions, the slope nearly equals zero, requiring lower power to retain force. This phenomenon has been overlooked and needs further analysis.
Finally, Table~\ref{tab:slope} shows the general CV's slope. CV’s slope tends to decrease as higher force is required according to before-fatigue conditions. The CV’s slope of 10\%MVC after-fatigue was first found, which has a contrasted trend, slightly increasing compared to other MVC levels. Note that this phenomenon has not been reported in any previous works. We suggest that this phenomenon is related to the reabsorption and distribution of ions such as H+ and K+ in the muscle membrane. This leads to an increase in CV and the shift of PSD to the right, as we mentioned earlier.

In summary, a novelty analysis on the PSD indicators was proposed, which generalized the compression trend of PSD at various MVC levels and conditions over time and evaluated the slope change of each parameter, such as MNF, RMS, and CV. In addition, the findings were presented according to 10\%MVC after-fatigue compared to other MVC levels.

\section{Conclusions}
In this study, we proposed a method that analyzes muscle fatigue indicators at various MVC levels to fully evaluate muscle conditions according to the before-fatigue, fatigue, and after-fatigue conditions, and analysis of the contraction time of each MVC level is also analyzed with the same indicators to understand better how indicators change to time. The decline of CV, MNF, an increase of RMS, and a shift of power spectrum to lower frequency according to different MVC levels before and during muscle fatigue all provide results that we expect about the trend of muscle fatigue. Regarding the significant increase in muscle contraction maintenance time at 10\%MVC level after muscle fatigue, we believe this increase in time is related to the reabsorption and redistribution of ions after muscle fatigue. As mentioned earlier, the change of CV has a solid relationship to the change of metabolism at the cell membrane. The CV also significantly impacts the amplitude and spectrum parameters, shown in the increased power spectrum density at higher frequencies and a slight increase in RMS and muscle force. However, this finding needs further investigation of the properties of the individual motor unit, such as firing rate, recruitment threshold, and synchronization.

\printbibliography
\end{document}